\documentstyle[emlines2]{article}

\def\hybrid{\topmargin 0pt      \oddsidemargin 0pt
        \headheight 0pt \headsep 0pt
        \textwidth 6.5in        % US paper
        \textheight 9.0in         % US paper
        \marginparwidth 0.0in
        \parskip 5pt plus 1pt   \jot = 1.5ex}
\catcode`\@=11
\def\marginnote#1{}

\newcount\hour
\newcount\minute
\newtoks\amorpm
\hour=\time\divide\hour by60
\minute=\time{\multiply\hour by60 \global\advance\minute by-\hour}
\edef\standardtime{{\ifnum\hour<12 \global\amorpm={am}%
        \else\global\amorpm={pm}\advance\hour by-12 \fi
        \ifnum\hour=0 \hour=12 \fi
        \number\hour:\ifnum\minute<10 0\fi\number\minute\the\amorpm}}
\edef\militarytime{\number\hour:\ifnum\minute<10 0\fi\number\minute}

\def\draftlabel#1{{\@bsphack\if@filesw {\let\thepage\relax
   \xdef\@gtempa{\write\@auxout{\string
      \newlabel{#1}{{\@currentlabel}{\thepage}}}}}\@gtempa
   \if@nobreak \ifvmode\nobreak\fi\fi\fi\@esphack}
        \gdef\@eqnlabel{#1}}
\def\@eqnlabel{}
\def\@vacuum{}
\def\draftmarginnote#1{\marginpar{\raggedright\scriptsize\tt#1}}

\def\draft{\oddsidemargin -.5truein
        \def\@oddfoot{\sl preliminary draft \hfil
        \rm\thepage\hfil\sl\today\quad\militarytime}
        \let\@evenfoot\@oddfoot \overfullrule 3pt
        \let\label=\draftlabel
        \let\marginnote=\draftmarginnote
   \def\@eqnnum{(\theequation)\rlap{\kern\marginparsep\tt\@eqnlabel}%
\global\let\@eqnlabel\@vacuum}  }
\def\titlepage{\@restonecolfalse\if@twocolumn\@restonecoltrue\onecolumn
     \else \newpage \fi \thispagestyle{empty}\c@page\z@
        \def\thefootnote{\fnsymbol{footnote}} }

\def\endtitlepage{\if@restonecol\twocolumn \else  \fi
        \def\thefootnote{\arabic{footnote}}
        \setcounter{footnote}{0}}  %\c@footnote\z@ }
\relax

\hybrid

\makeatletter
\newdimen\normalarrayskip              % skip between lines
\newdimen\minarrayskip                 % minimal skip between lines
\normalarrayskip\baselineskip
\minarrayskip\jot
\newif\ifold             \oldtrue            \def\new{\oldfalse}
\def\arraymode{\ifold\relax\else\displaystyle\fi} % mode of array enrties
\def\eqnumphantom{\phantom{(\theequation)}}     % right phantom in eqnarray
\def\@arrayskip{\ifold\baselineskip\z@\lineskip\z@
     \else
     \baselineskip\minarrayskip\lineskip2\minarrayskip\fi}

\def\@arrayclassz{\ifcase \@lastchclass \@acolampacol \or
\@ampacol \or \or \or \@addamp \or
   \@acolampacol \or \@firstampfalse \@acol \fi
\edef\@preamble{\@preamble
  \ifcase \@chnum
     \hfil$\relax\arraymode\@sharp$\hfil
     \or $\relax\arraymode\@sharp$\hfil
     \or \hfil$\relax\arraymode\@sharp$\fi}}

\def\@array[#1]#2{\setbox\@arstrutbox=\hbox{\vrule
     height\arraystretch \ht\strutbox
     depth\arraystretch \dp\strutbox
     width\z@}\@mkpream{#2}\edef\@preamble{\halign \noexpand\@halignto
\bgroup \tabskip\z@ \@arstrut \@preamble \tabskip\z@ \cr}%
\let\@startpbox\@@startpbox \let\@endpbox\@@endpbox
  \if #1t\vtop \else \if#1b\vbox \else \vcenter \fi\fi
  \bgroup \let\par\relax
  \let\@sharp##\let\protect\relax
  \@arrayskip\@preamble}
\def\eqnarray{\stepcounter{equation}%
              \let\@currentlabel=\theequation
              \global\@eqnswtrue
              \global\@eqcnt\z@
              \tabskip\@centering
              \let\\=\@eqncr
              $$%
 \halign to \displaywidth\bgroup
    \eqnumphantom\@eqnsel\hskip\@centering
    $\displaystyle \tabskip\z@ {##}$%
    &\global\@eqcnt\@ne \hskip 2\arraycolsep
         $\displaystyle\arraymode{##}$\hfil
    &\global\@eqcnt\tw@ \hskip 2\arraycolsep
         $\displaystyle\tabskip\z@{##}$\hfil
         \tabskip\@centering
    &{##}\tabskip\z@\cr}
\makeatother

\begingroup\ifx\undefined\newsymbol \else\def\input#1 {\endgroup}\fi
\input amssym.def \relax
\input amssym
\newfont{\hr}{msbm10}
\newfont{\ams}{msam10}
\def\bea{\begin{eqnarray}}
\def\eea{\end{eqnarray}}
\def\nn{\nonumber}

\def\beq{\begin{equation}}
\def\eeq{\end{equation}}
\def\be{\beq\new\begin{array}{c}}
\def\ee{\end{array}\eeq}
\def\Tr{{\rm Tr}}
\def\res{{\rm res}}
\def\Bf#1{\mbox{\boldmath $#1$}}
\def\balpha{{\Bf\alpha}}

\def\bphi{{\Bf\phi}}

\def\Im{{\rm Im}}
\def\Re{{\rm Re}}

\def\rank{{\rm rank}}
\def\2{{1\over 2}}
\def\N2{${\cal N}=2$}
\begin{document}

%\draft                               %SWITCH ON/OFF DRAFT VERSION%

\begin{titlepage}
\setcounter{footnote}0
\begin{center}
\hfill FIAN/TD-12/97\\
\hfill ITEP/TH-39/97\\
\hfill hep-th/9709001\\
%\hfill{\it preliminary draft}\\
\vspace{0.3in}
{\LARGE\bf Seiberg-Witten Theory, Integrable Systems and D-branes
\footnote{based on the talks given at NATO Advanced Research Workshop on
Theoretical Physics "New Developments in Quantum Field Theory",
Zakopane, 14-20 June 1997 and IV International Conference "Conformal Field
Theories and Integrable Models", Chernogolovka, 23-27 June 1997.}}
\\
\bigskip
\bigskip
\bigskip
{\Large A.Marshakov
\footnote{E-mail address: mars@lpi.ac.ru, andrei@heron.itep.ru,
marshakov@nbivms.nbi.dk}}
\\
\bigskip
{\it Theory Department,  P. N. Lebedev Physics
Institute , Leninsky prospect, 53, Moscow,~117924, Russia\\
and ITEP, Moscow 117259, Russia}
\end{center}
\bigskip \bigskip

\begin{abstract}
In this note it is demonstrated how the Seiberg-Witten solutions and related
integrable systems may arise from certain brane configurations in $M$-theory.
Some subtleties of the formulation of the Seiberg-Witten theory via
integrable systems are discussed and interpreted along the lines of general
picture os string/$M$-theory dualities.
\end{abstract}

\end{titlepage}

%\newpage
%\tableofcontents
\newpage
\setcounter{footnote}0
%\footnotesize

\section{Introduction}

A lot of ideas has appeared recently in theoretical physics due to
developments in what can be called nonperturbative string theory or
$M$-theory
\footnote{There is no fixed terminology yet in this field
-- sometimes the term $M$-theory is applied in more "narrow" sense -- to
the theory of membranes, M(atrix) models etc. In this note we will use the
term $M$-theory in wide sense -- identify
it with the hypothetical (11-dimensional) nonperturbative "string"
theory.}. The basic concept is that the physically interesting quantum
field theories (QFT's) could be considered as various vacua of $M$-theory
and the stringy symmetries or dualities may relate spectra and
correlation functions in one QFT with those in another QFT; then typical
string duality $R \leftrightarrow {\alpha '\over R}$ allows in principle
to relate
perturbative regime in one model with the nonperturbative in another.

This general idea at the moment was put to more solid ground only for
the case of complex backgrounds in string theory ($\equiv$ supersymmetric
(SUSY) quantum field theories). In such theories physical data of the
model (masses, couplings, etc) can be considered as functions on {\em
moduli spaces} of complex manifolds and the duality symmetry can be regarded
as action of a modular group. Useful information can be extracted by
powerful machinery of complex geometry. Despite some progress achieved
along these lines in the popular scheme of string compactifications on
the Calabi-Yau manifolds (see, for example \cite{CaYa} and references
therein), rigorous statements about the net result of
nonperturbative string theory can be made till now only when nontrivial
complex geometry can be effectively reduced to the geometry of
one-dimensional complex ($\equiv $ two-dimensional real) manifolds
-- complex curves $\Sigma $. One of the most interesting examples of
exact statements in this field is the Seiberg-Witten anzatz for the
Coulomb phase of \N2 SUSY Yang-Mills theories in four dimensions
\cite{SWetc}.

In this note I will explain once more how the Seiberg-Witten (SW) anzatz
arises from the brane configurations in $M$-theory along the lines
of \cite{Witten,MMM}. Following the approach of \cite{GKMMM,after,M},
the language of integrable systems will be used for the formulations of the
exact results in \N2 SUSY four-dimensional gauge theories. I will try
to pay attention to the subtleties of the exact formulation of the
results in these terms and demonstrate how some of them are governed by the
Diaconescu-Hanany-Witten-Witten (DHWW) construction.

\section{Seiberg-Witten Anzatz: Integrable Systems}

For the \N2 SUSY gauge theory the SW anzatz can be
{\em formulated} in the following way.
The \N2 SUSY vector multiplet has necessarily (complex) scalars
with the potential $V(\bphi )=\Tr [\bphi, \bphi ^{\dagger}]^2$ whose
minima (after factorization over the
gauge group) correspond to the diagonal ($[\bphi, \bphi ^{\dagger}] = 0$),
constant and (in the theory with $SU(N_c)$ gauge group) traceless matrices.
Their invariants
\be\label{polyn}
\det (\lambda - \bphi) \equiv P_{N_c}(\lambda ) = \sum_{k=0}^{N_c}
s_{N_c-k}\lambda ^k
\ee
(the total number of algebraically independent ones is
$\rank\ SU(N_c) = N_c -1$) parameterize the moduli space of the theory.
Due to the Higgs effect the off-diagonal part of the gauge field
${\bf A}_{\mu}$ becomes massive, since
\be\label{comm}
[\bphi , {\bf A}_{\mu}]_{ij} = (\phi_i-\phi_j){\bf A}_{\mu}^{ij}
\ee
while the diagonal part, as it follows from (\ref{comm}) remains massless,
i.e. the gauge group $G = SU(N_c)$ breaks down to
$U(1)^{\rank G} = U(1) ^{N_c -1}$.

The effective abelian theory is formulated in terms of a
finite-dimensional integrable system: the spectral curve $\Sigma $
defined over the genus-dimensional subspace of the full moduli space, e.g.
\be\label{suncu}
\Lambda^{N_c}\left(w + \frac{1}{w}\right) = 2P_{N_c}(\lambda)
\ee
for the pure $SU(N_c)$ gauge theory
\footnote{The genus of the curve (\ref{suncu})
is $N_c-1$, i.e. exactly equal to the number of independent parameters of
the polynomial $P_{N_c}(\lambda )$ (\ref{polyn}).}; and the generating
differential
\be\label{dS}
dS = \lambda{dw\over w}
\ee
whose basic property is that its derivatives over $N_c-1$ moduli give rise
to holomorphic differentials. The data $(\Sigma, dS)$ with such properties
are exactly the definition of the integrable system in the sense of
\cite{DKN} (see \cite{M} and references therein for details). The period
matrix of $\Sigma $ $T_{ij}({\bf a})$ as a function of the action
variables
\be
{\bf a} = \oint _{\bf A}dS \ \ \ \ \ \ \ \ {\bf a}^D = \oint _{\bf B}dS
\nn \\
T_{ij} = {\partial a^D_i\over\partial a_j}
\ee
gives the set of coupling constants in the effective abelian
$U(1)^{N_c-1}$ theory while action variables themselves are identified
with the masses of the BPS states $M^2 \sim |{\bf na} + {\bf ma}_D|^2$ with
the $({\bf n},{\bf m})$ "electric" and "magnetic" charges.

\section{Spectral Curve as Topologically Nontrivial Part of
M-theory 5-brane World Volume}

Let us show now that the spectral curve and generating differential
can naturally arise from brane configurations \cite{Witten}.
\begin{itemize}
\item First step is to obtain a gauge group $SU(N_c)$ broken down to
$U(1)^{N_c-1}$. The most elegant way of doing this in string theory is
to introduce D-branes into type II string theory -- the submanifolds in
target space where strings can have their ends. $N_c$ parallel D-branes
would correspond exactly to what we need now since string strechted
between $i$-th and $j$-th brane ($i,j=1,\dots,N_c$) (see Fig.1)
\begin{figure}
%TexCad Options
%\grade{\on}
%\emlines{\on}
%\beziermacro{\off}
%\reduce{\on}
%\snapping{\off}
%\quality{2.00}
%\graddiff{0.01}
%\snapasp{1}
%\zoom{7.23}
\special{em:linewidth 0.4pt}
\unitlength 1.00mm
\linethickness{0.4pt}
\begin{picture}(114.67,100.33)
\emline{60.00}{100.33}{1}{110.00}{100.33}{2}
\emline{60.00}{100.33}{1}{110.00}{100.33}{2}
\emline{60.00}{92.00}{3}{110.00}{92.00}{4}
\emline{60.00}{83.67}{5}{110.00}{83.67}{6}
\emline{60.00}{75.00}{7}{110.00}{75.00}{8}
\put(85.00,71.67){\circle*{0.67}}
\put(85.00,69.00){\circle*{0.67}}
\put(85.00,66.33){\circle*{0.67}}
\emline{60.00}{62.00}{9}{110.00}{62.00}{10}
\put(114.67,92.00){\makebox(0,0)[cc]{$i$}}
\put(114.33,75.33){\makebox(0,0)[cc]{$j$}}
\put(94.94,92.00){\circle*{0.67}}
\put(93.85,74.99){\circle*{0.67}}
\emline{95.00}{92.02}{11}{93.99}{91.00}{12}
\emline{93.99}{91.00}{13}{95.00}{89.99}{14}
\emline{95.00}{89.99}{15}{95.97}{89.02}{16}
\emline{95.97}{89.02}{17}{94.96}{88.00}{18}
\emline{94.96}{88.00}{19}{93.94}{86.99}{20}
\emline{94.91}{83.99}{21}{93.90}{82.98}{22}
\emline{94.87}{79.98}{23}{93.85}{78.96}{24}
\emline{94.82}{75.97}{25}{93.81}{74.95}{26}
\emline{93.94}{86.99}{27}{94.96}{85.97}{28}
\emline{93.90}{82.98}{29}{94.91}{81.96}{30}
\emline{93.85}{78.96}{31}{94.87}{77.95}{32}
\emline{94.96}{85.97}{33}{95.93}{85.01}{34}
\emline{94.91}{81.96}{35}{95.88}{80.99}{36}
\emline{94.87}{77.95}{37}{95.83}{76.98}{38}
\emline{95.93}{85.01}{39}{94.91}{83.99}{40}
\emline{95.88}{80.99}{41}{94.87}{79.98}{42}
\emline{95.83}{76.98}{43}{94.82}{75.97}{44}
\end{picture}
\caption{
Open strings, stretched between $D$-branes induce the interaction via
non-Abelian gauge fields ${\bf A}^{ij}$.}
\end{figure}
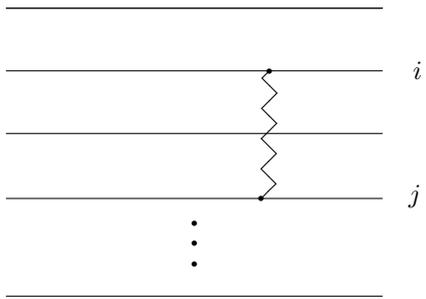
will have a
vector
field ${\bf A}^{ij}$ in its spectrum such that mass of this vector field
is proportional to the length of the string, i.e. to the distance between
$i$-th and $j$-th branes. This the $U(1)^{N_c-1}$ massless factor will come
from strings having both ends on the same D-brane while the ${\bf A}^{ij}$
fields with $i\neq j$ will acquire "Higgs" masses (\ref{comm}) where
scalars vev's are us usual proportional to the "transverse" co-ordinate
of the D-brane $\phi \sim {\sqrt{{\vec x}_{\bot}^2}\over\alpha '}$.

\item Next step is that from 10-dimensional type II string theory
(${\bf A} = \|{\bf A}^{ij}\| $ is 10-dimensional gauge field in string
picture) one wants to
get 4-dimensional one. A natural way to reduce the number of space-time
dimensions is to restrict ourselves to the effective theory on D-brane
world volume. The world volume of the Dirichlet $p$-brane is
$p+1$-dimensional, so naively in order to get 4-dimensional theory one
should consider D3-branes. This scenario is quite possible and realized
in another context; however to get the SW anzatz it is better to use
another option, the DHWW brane configuration with the $N_c$ parallel
D4-branes streched between two vertical walls (see Fig.2),
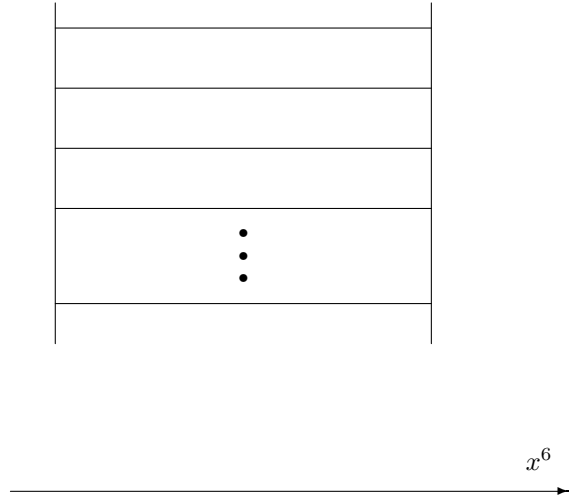
\begin{figure}
%TexCad Options
%\grade{\on}
%\emlines{\on}
%\beziermacro{\off}
%\reduce{\on}
%\snapping{\off}
%\quality{2.00}
%\graddiff{0.01}
%\snapasp{1}
%\zoom{1.00}
\special{em:linewidth 0.4pt}
\unitlength 1.00mm
\linethickness{0.4pt}
\begin{picture}(118.00,103.33)(-10.00,0.00)
\emline{50.00}{100.00}{1}{100.00}{100.00}{2}
\emline{50.00}{92.00}{3}{100.00}{92.00}{4}
\emline{50.00}{84.00}{5}{100.00}{84.00}{6}
\emline{50.00}{76.00}{7}{100.00}{76.00}{8}
\put(75.00,72.67){\circle*{0.94}}
\put(75.00,69.67){\circle*{0.94}}
\put(75.00,66.67){\circle*{0.94}}
\emline{50.00}{63.33}{9}{100.00}{63.33}{10}
\emline{50.00}{103.33}{11}{50.00}{58.00}{12}
\emline{100.00}{103.33}{13}{100.00}{58.00}{14}
%\vector(44.00,38.33)(118.00,38.33)
\put(118.00,38.33){\vector(1,0){0.2}}
\emline{44.00}{38.33}{15}{118.00}{38.33}{16}
%\end
\put(114.33,42.67){\makebox(0,0)[cc]{$x^6$}}
\end{picture}
\caption{The 4-branes restricted by 5-branes to the finite volume
(in horisontal $x^6$-direction) give rise to macroscopically 4-dimensional
theory.}
\end{figure}
so that the
naive
5-dimensional D4-world-volume theory is macroscopically (in the light
sector) 4-dimensional by conventional Kaluza-Klein argument for a
system compactified on a circle or put into a box.

The role of vertical walls should be played by 5-branes \cite{Witten}, this
follows from the $\beta $-function considerations: the logariphmic behaivior
of the macroscopic coupling constant can be ensured in the first
approximation if corresponding co-ordinate ($x^6$) has logariphmic
behaivior as a function of "transverse" direction, i.e. satisfy
{\em two}-dimensional Laplace equation. The effective space is
two-dimensional if parallel D-branes are streched between the 5-branes.

\item The obtained picture of 4 and 5 branes in 10 dimensions is of course
very rough and true in (semi)\-classical approximation. In particular it is
naively singular at the points where 4-branes meet 5-branes. These
singularities were resolved by Witten \cite{Witten} who suggested to
put the whole picture into 11-dimensional target space of $M$-theory
with compact 11-th dimension and to consider D4-branes as $M$-theory
5-branes compactified to 11-th dimension. Thus the picture in Fig.2
becomes similiar rather to the surface of "swedish ladder"
\footnote{I am grateful to V.Kazakov for this not quite exact but
illuminating comparison.} and apart of macroscopic directions
$x^0,\dots,x^3$ looks like (non-compact) Riemann surface with rather
special properties (see Fig.3).
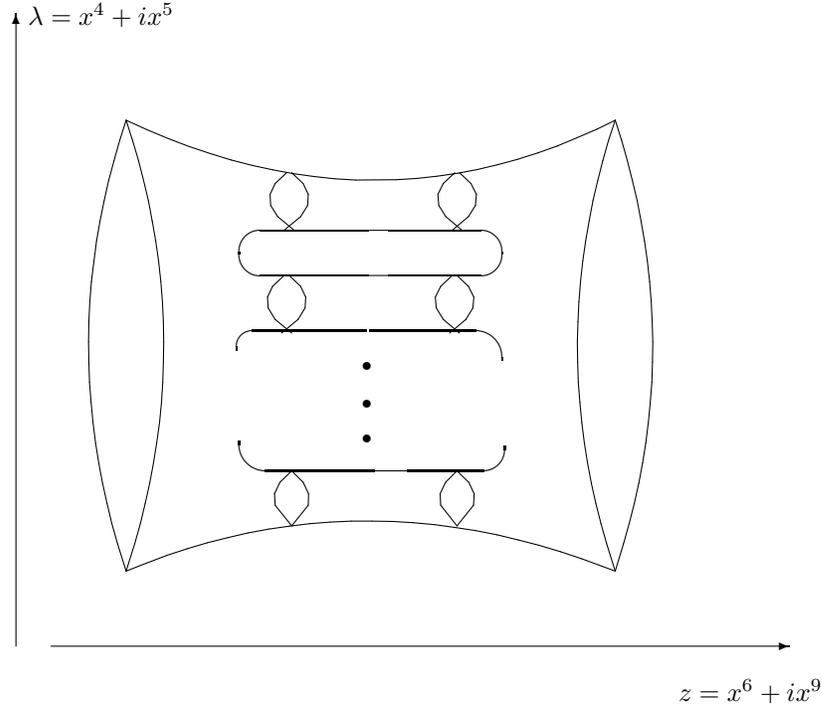
\begin{figure}
%TexCad Options
%\grade{\on}
%\emlines{\on}
%\beziermacro{\off}
%\reduce{\on}
%\snapping{\off}
%\quality{2.00}
%\graddiff{0.01}
%\snapasp{1}
%\zoom{1.00}
\special{em:linewidth 0.4pt}
\unitlength 1mm
\linethickness{0.4pt}
\begin{picture}(128.00,124.00)(-10.00,0.00)
%\bezier{252}(40.00,110.00)(30.00,80.00)(40.00,50.00)
\emline{40.00}{110.00}{1}{39.15}{107.32}{2}
\emline{39.15}{107.32}{3}{38.37}{104.63}{4}
\emline{38.37}{104.63}{5}{37.68}{101.95}{6}
\emline{37.68}{101.95}{7}{37.06}{99.27}{8}
\emline{37.06}{99.27}{9}{36.53}{96.58}{10}
\emline{36.53}{96.58}{11}{36.07}{93.90}{12}
\emline{36.07}{93.90}{13}{35.70}{91.22}{14}
\emline{35.70}{91.22}{15}{35.40}{88.53}{16}
\emline{35.40}{88.53}{17}{35.19}{85.85}{18}
\emline{35.19}{85.85}{19}{35.06}{83.17}{20}
\emline{35.06}{83.17}{21}{35.00}{80.48}{22}
\emline{35.00}{80.48}{23}{35.03}{77.80}{24}
\emline{35.03}{77.80}{25}{35.13}{75.12}{26}
\emline{35.13}{75.12}{27}{35.32}{72.44}{28}
\emline{35.32}{72.44}{29}{35.58}{69.75}{30}
\emline{35.58}{69.75}{31}{35.93}{67.07}{32}
\emline{35.93}{67.07}{33}{36.35}{64.39}{34}
\emline{36.35}{64.39}{35}{36.86}{61.70}{36}
\emline{36.86}{61.70}{37}{37.45}{59.02}{38}
\emline{37.45}{59.02}{39}{38.11}{56.34}{40}
\emline{38.11}{56.34}{41}{38.86}{53.65}{42}
\emline{38.86}{53.65}{43}{40.00}{50.00}{44}
%\end
%\bezier{252}(40.00,110.00)(50.00,80.00)(40.00,50.00)
\emline{40.00}{110.00}{45}{40.85}{107.32}{46}
\emline{40.85}{107.32}{47}{41.63}{104.63}{48}
\emline{41.63}{104.63}{49}{42.32}{101.95}{50}
\emline{42.32}{101.95}{51}{42.94}{99.27}{52}
\emline{42.94}{99.27}{53}{43.47}{96.58}{54}
\emline{43.47}{96.58}{55}{43.93}{93.90}{56}
\emline{43.93}{93.90}{57}{44.30}{91.22}{58}
\emline{44.30}{91.22}{59}{44.60}{88.53}{60}
\emline{44.60}{88.53}{61}{44.81}{85.85}{62}
\emline{44.81}{85.85}{63}{44.94}{83.17}{64}
\emline{44.94}{83.17}{65}{45.00}{80.48}{66}
\emline{45.00}{80.48}{67}{44.97}{77.80}{68}
\emline{44.97}{77.80}{69}{44.87}{75.12}{70}
\emline{44.87}{75.12}{71}{44.68}{72.44}{72}
\emline{44.68}{72.44}{73}{44.42}{69.75}{74}
\emline{44.42}{69.75}{75}{44.07}{67.07}{76}
\emline{44.07}{67.07}{77}{43.65}{64.39}{78}
\emline{43.65}{64.39}{79}{43.14}{61.70}{80}
\emline{43.14}{61.70}{81}{42.55}{59.02}{82}
\emline{42.55}{59.02}{83}{41.89}{56.34}{84}
\emline{41.89}{56.34}{85}{41.14}{53.65}{86}
\emline{41.14}{53.65}{87}{40.00}{50.00}{88}
%\end
%\bezier{252}(105.00,110.00)(95.00,80.00)(105.00,50.00)
\emline{105.00}{110.00}{89}{104.15}{107.32}{90}
\emline{104.15}{107.32}{91}{103.37}{104.63}{92}
\emline{103.37}{104.63}{93}{102.68}{101.95}{94}
\emline{102.68}{101.95}{95}{102.06}{99.27}{96}
\emline{102.06}{99.27}{97}{101.53}{96.58}{98}
\emline{101.53}{96.58}{99}{101.07}{93.90}{100}
\emline{101.07}{93.90}{101}{100.70}{91.22}{102}
\emline{100.70}{91.22}{103}{100.40}{88.53}{104}
\emline{100.40}{88.53}{105}{100.19}{85.85}{106}
\emline{100.19}{85.85}{107}{100.06}{83.17}{108}
\emline{100.06}{83.17}{109}{100.00}{80.48}{110}
\emline{100.00}{80.48}{111}{100.03}{77.80}{112}
\emline{100.03}{77.80}{113}{100.13}{75.12}{114}
\emline{100.13}{75.12}{115}{100.32}{72.44}{116}
\emline{100.32}{72.44}{117}{100.58}{69.75}{118}
\emline{100.58}{69.75}{119}{100.93}{67.07}{120}
\emline{100.93}{67.07}{121}{101.35}{64.39}{122}
\emline{101.35}{64.39}{123}{101.86}{61.70}{124}
\emline{101.86}{61.70}{125}{102.45}{59.02}{126}
\emline{102.45}{59.02}{127}{103.11}{56.34}{128}
\emline{103.11}{56.34}{129}{103.86}{53.65}{130}
\emline{103.86}{53.65}{131}{105.00}{50.00}{132}
%\end
%\bezier{252}(105.00,110.00)(115.00,80.00)(105.00,50.00)
\emline{105.00}{110.00}{133}{105.85}{107.32}{134}
\emline{105.85}{107.32}{135}{106.63}{104.63}{136}
\emline{106.63}{104.63}{137}{107.32}{101.95}{138}
\emline{107.32}{101.95}{139}{107.94}{99.27}{140}
\emline{107.94}{99.27}{141}{108.47}{96.58}{142}
\emline{108.47}{96.58}{143}{108.93}{93.90}{144}
\emline{108.93}{93.90}{145}{109.30}{91.22}{146}
\emline{109.30}{91.22}{147}{109.60}{88.53}{148}
\emline{109.60}{88.53}{149}{109.81}{85.85}{150}
\emline{109.81}{85.85}{151}{109.94}{83.17}{152}
\emline{109.94}{83.17}{153}{110.00}{80.48}{154}
\emline{110.00}{80.48}{155}{109.97}{77.80}{156}
\emline{109.97}{77.80}{157}{109.87}{75.12}{158}
\emline{109.87}{75.12}{159}{109.68}{72.44}{160}
\emline{109.68}{72.44}{161}{109.42}{69.75}{162}
\emline{109.42}{69.75}{163}{109.07}{67.07}{164}
\emline{109.07}{67.07}{165}{108.65}{64.39}{166}
\emline{108.65}{64.39}{167}{108.14}{61.70}{168}
\emline{108.14}{61.70}{169}{107.55}{59.02}{170}
\emline{107.55}{59.02}{171}{106.89}{56.34}{172}
\emline{106.89}{56.34}{173}{106.14}{53.65}{174}
\emline{106.14}{53.65}{175}{105.00}{50.00}{176}
%\end
%\bezier{288}(40.00,110.00)(73.67,94.00)(105.00,110.00)
\emline{40.00}{110.00}{177}{42.81}{108.72}{178}
\emline{42.81}{108.72}{179}{45.61}{107.55}{180}
\emline{45.61}{107.55}{181}{48.41}{106.49}{182}
\emline{48.41}{106.49}{183}{51.20}{105.54}{184}
\emline{51.20}{105.54}{185}{53.97}{104.71}{186}
\emline{53.97}{104.71}{187}{56.74}{103.99}{188}
\emline{56.74}{103.99}{189}{59.51}{103.38}{190}
\emline{59.51}{103.38}{191}{62.26}{102.88}{192}
\emline{62.26}{102.88}{193}{65.01}{102.49}{194}
\emline{65.01}{102.49}{195}{67.74}{102.21}{196}
\emline{67.74}{102.21}{197}{70.47}{102.05}{198}
\emline{70.47}{102.05}{199}{75.91}{102.06}{200}
\emline{75.91}{102.06}{201}{78.61}{102.23}{202}
\emline{78.61}{102.23}{203}{81.31}{102.52}{204}
\emline{81.31}{102.52}{205}{84.00}{102.91}{206}
\emline{84.00}{102.91}{207}{86.68}{103.42}{208}
\emline{86.68}{103.42}{209}{89.35}{104.04}{210}
\emline{89.35}{104.04}{211}{92.02}{104.77}{212}
\emline{92.02}{104.77}{213}{94.67}{105.62}{214}
\emline{94.67}{105.62}{215}{97.32}{106.57}{216}
\emline{97.32}{106.57}{217}{99.96}{107.64}{218}
\emline{99.96}{107.64}{219}{102.59}{108.82}{220}
\emline{102.59}{108.82}{221}{105.00}{110.00}{222}
%\end
%\bezier{280}(40.00,50.00)(71.67,63.33)(105.00,50.00)
\emline{40.00}{50.00}{223}{42.69}{51.08}{224}
\emline{42.69}{51.08}{225}{45.38}{52.07}{226}
\emline{45.38}{52.07}{227}{48.08}{52.96}{228}
\emline{48.08}{52.96}{229}{50.79}{53.76}{230}
\emline{50.79}{53.76}{231}{53.50}{54.46}{232}
\emline{53.50}{54.46}{233}{56.22}{55.06}{234}
\emline{56.22}{55.06}{235}{58.95}{55.57}{236}
\emline{58.95}{55.57}{237}{61.68}{55.98}{238}
\emline{61.68}{55.98}{239}{64.42}{56.29}{240}
\emline{64.42}{56.29}{241}{67.16}{56.51}{242}
\emline{67.16}{56.51}{243}{69.91}{56.64}{244}
\emline{69.91}{56.64}{245}{72.66}{56.66}{246}
\emline{72.66}{56.66}{247}{75.42}{56.60}{248}
\emline{75.42}{56.60}{249}{78.19}{56.43}{250}
\emline{78.19}{56.43}{251}{80.96}{56.17}{252}
\emline{80.96}{56.17}{253}{83.74}{55.82}{254}
\emline{83.74}{55.82}{255}{86.53}{55.36}{256}
\emline{86.53}{55.36}{257}{89.32}{54.82}{258}
\emline{89.32}{54.82}{259}{92.11}{54.17}{260}
\emline{92.11}{54.17}{261}{94.92}{53.43}{262}
\emline{94.92}{53.43}{263}{97.72}{52.60}{264}
\emline{97.72}{52.60}{265}{100.54}{51.67}{266}
\emline{100.54}{51.67}{267}{105.00}{50.00}{268}
%\end
\put(72.17,92.33){\oval(34.33,6.00)[l]}
\put(74.83,92.33){\oval(30.33,6.00)[r]}
\put(72.00,79.33){\oval(34.67,5.33)[lt]}
\put(73.00,67.33){\oval(36.00,8.00)[lb]}
\put(77.33,66.67){\oval(26.00,6.67)[rb]}
\put(72.33,78.00){\oval(35.33,8.00)[rt]}
%\bezier{52}(61.33,103.00)(56.33,99.33)(62.33,95.33)
\emline{61.33}{103.00}{269}{59.82}{101.58}{270}
\emline{59.82}{101.58}{271}{59.11}{100.13}{272}
\emline{59.11}{100.13}{273}{59.23}{98.66}{274}
\emline{59.23}{98.66}{275}{60.15}{97.16}{276}
\emline{60.15}{97.16}{277}{62.33}{95.33}{278}
%\end
%\bezier{52}(62.00,103.00)(67.00,99.33)(61.00,95.33)
\emline{62.00}{103.00}{279}{63.52}{101.58}{280}
\emline{63.52}{101.58}{281}{64.22}{100.13}{282}
\emline{64.22}{100.13}{283}{64.11}{98.66}{284}
\emline{64.11}{98.66}{285}{63.18}{97.16}{286}
\emline{63.18}{97.16}{287}{61.00}{95.33}{288}
%\end
%\bezier{52}(83.67,103.00)(78.67,99.33)(84.67,95.33)
\emline{83.67}{103.00}{289}{82.15}{101.58}{290}
\emline{82.15}{101.58}{291}{81.45}{100.13}{292}
\emline{81.45}{100.13}{293}{81.56}{98.66}{294}
\emline{81.56}{98.66}{295}{82.48}{97.16}{296}
\emline{82.48}{97.16}{297}{84.67}{95.33}{298}
%\end
%\bezier{52}(84.33,103.00)(89.33,99.33)(83.33,95.33)
\emline{84.33}{103.00}{299}{85.85}{101.58}{300}
\emline{85.85}{101.58}{301}{86.55}{100.13}{302}
\emline{86.55}{100.13}{303}{86.44}{98.66}{304}
\emline{86.44}{98.66}{305}{85.52}{97.16}{306}
\emline{85.52}{97.16}{307}{83.33}{95.33}{308}
%\end
%\bezier{52}(61.00,89.33)(56.00,85.67)(62.00,81.67)
\emline{61.00}{89.33}{309}{59.48}{87.91}{310}
\emline{59.48}{87.91}{311}{58.78}{86.46}{312}
\emline{58.78}{86.46}{313}{58.89}{84.99}{314}
\emline{58.89}{84.99}{315}{59.82}{83.50}{316}
\emline{59.82}{83.50}{317}{62.00}{81.67}{318}
%\end
%\bezier{52}(61.67,89.33)(66.67,85.67)(60.67,81.67)
\emline{61.67}{89.33}{319}{63.18}{87.91}{320}
\emline{63.18}{87.91}{321}{63.89}{86.46}{322}
\emline{63.89}{86.46}{323}{63.77}{84.99}{324}
\emline{63.77}{84.99}{325}{62.85}{83.50}{326}
\emline{62.85}{83.50}{327}{60.67}{81.67}{328}
%\end
%\bezier{52}(83.33,89.33)(78.33,85.67)(84.33,81.67)
\emline{83.33}{89.33}{329}{81.82}{87.91}{330}
\emline{81.82}{87.91}{331}{81.11}{86.46}{332}
\emline{81.11}{86.46}{333}{81.23}{84.99}{334}
\emline{81.23}{84.99}{335}{82.15}{83.50}{336}
\emline{82.15}{83.50}{337}{84.33}{81.67}{338}
%\end
%\bezier{52}(84.00,89.33)(89.00,85.67)(83.00,81.67)
\emline{84.00}{89.33}{339}{85.52}{87.91}{340}
\emline{85.52}{87.91}{341}{86.22}{86.46}{342}
\emline{86.22}{86.46}{343}{86.11}{84.99}{344}
\emline{86.11}{84.99}{345}{85.18}{83.50}{346}
\emline{85.18}{83.50}{347}{83.00}{81.67}{348}
%\end
\emline{71.00}{95.33}{349}{76.33}{95.33}{350}
\emline{70.33}{89.33}{351}{76.33}{89.33}{352}
\emline{70.00}{63.33}{353}{79.67}{63.33}{354}
%\bezier{48}(62.00,63.33)(57.33,59.67)(62.00,56.00)
\emline{62.00}{63.33}{355}{60.46}{61.81}{356}
\emline{60.46}{61.81}{357}{59.73}{60.28}{358}
\emline{59.73}{60.28}{359}{59.81}{58.75}{360}
\emline{59.81}{58.75}{361}{62.00}{56.00}{362}
%\end
%\bezier{48}(62.00,63.33)(66.67,59.67)(62.00,56.00)
\emline{62.00}{63.33}{363}{63.54}{61.81}{364}
\emline{63.54}{61.81}{365}{64.27}{60.28}{366}
\emline{64.27}{60.28}{367}{64.19}{58.75}{368}
\emline{64.19}{58.75}{369}{62.00}{56.00}{370}
%\end
%\bezier{48}(84.00,63.33)(79.33,59.67)(84.00,56.00)
\emline{84.00}{63.33}{371}{82.46}{61.81}{372}
\emline{82.46}{61.81}{373}{81.73}{60.28}{374}
\emline{81.73}{60.28}{375}{81.81}{58.75}{376}
\emline{81.81}{58.75}{377}{84.00}{56.00}{378}
%\end
%\bezier{48}(84.00,63.33)(88.67,59.67)(84.00,56.00)
\emline{84.00}{63.33}{379}{85.54}{61.81}{380}
\emline{85.54}{61.81}{381}{86.27}{60.28}{382}
\emline{86.27}{60.28}{383}{86.19}{58.75}{384}
\emline{86.19}{58.75}{385}{84.00}{56.00}{386}
%\end
\put(72.00,77.33){\circle*{0.94}}
\put(72.00,72.33){\circle*{0.94}}
\put(72.00,67.67){\circle*{0.94}}
%\vector(30.00,40.00)(128.00,40.00)
\put(128.00,40.00){\vector(1,0){0.2}}
\emline{30.00}{40.00}{387}{128.00}{40.00}{388}
%\end
%\vector(25.33,40.00)(25.33,124.00)
\put(25.33,124.00){\vector(0,1){0.2}}
\emline{25.33}{40.00}{389}{25.33}{124.00}{390}
%\end
\put(123.00,34.00){\makebox(0,0)[cc]{$z=x^6+ix^9$}}
\put(36.67,124.00){\makebox(0,0)[cc]{$\lambda = x^4+ix^5$}}
\end{picture}
\caption{The brane configuration, represented as a result of
"blowing up" Fig.2 -- the ladder turns into hyperelliptic
Riemann surface being at the same time
$N_c$-fold covering of the horisontal cylinder.}
\end{figure}
\item In other words, one gets a $5$-brane parameterized by
$(x^0,x^1,x^2,x^3,x^6,x^{9})$, which
leaving aside four flat dimensions ($x^0,x^1,x^2,x^3$) along these lines
ends up to $N_c$ cylinders $R\times S^1$ embedded
into the target space along, say, $(x^6,x^9)$ dimensions
(using notation $z = x^6+ix^9$ for the
corresponding complex co-ordinate).
Different cylinders have different positions in
the space $V^{\bot} = (x^4,x^5,x^7,x^8)$.
Moreover the cylinders are all glued together (see Fig.3).
The "effective" two-dimensional subspace of $V^{\bot}$
we will describe it in terms of the complex coordinate $\lambda = x^4 +ix^5$.

Introducing coordinate $w = e^{z}$ to describe a cylinder,
we see that the system of non-interacting branes (Fig.1) is
given by $z$-independent equation
\be
P_{N_c}(\lambda) = \prod_{\alpha = 1}^{N_c}
(\lambda - \lambda_\alpha) = 0,
\label{nibr}
\ee
while their bound state (Fig.3) is described rather by the complex curve
$\Sigma _{N_c}$ (\ref{suncu}) or:
\be
\Lambda^{N_c}\cosh z = P_{N_c}(\lambda)
\label{todacur}
\ee
In the weak-coupling limit  $\Lambda \rightarrow 0$
(i.e. $\frac{1}{g^2} \sim \log\Lambda \rightarrow \infty$)
one comes back to  disjoint branes (\ref{nibr})
\footnote{Eqs.(\ref{suncu}), (\ref{todacur})
and Fig.3 decribe a hyperelliptic curve --
a double covering of a punctured Riemann sphere,
\be
y^2 = \frac{\Lambda^{2N_c}}{4}\left(w - \frac{1}{w}\right)^2 =
P_{N_c}^2(\lambda) - \Lambda^{2N_c}
\nn
\ee
}.
Thus we finally got a $5$-brane of topology
$R^3\times\Sigma _{N_c}$ embedded into a subspace $R^6\times S^1$
(spanned by $x^1,...,x^6,x^9$) of the full target space. The
periodic coordinate is
\be\label{perco}
x^9 = \arg P_{N_c}(\lambda) = \Im\log P_{N_c}(\lambda) =
\sum _{\alpha = 1}^{N_c} \arg (\lambda - \lambda _{\alpha})
\ee

\end{itemize}

\section{Integrable equations from brane picture}

The arguments of the previous section show that the nontrivial part
of the 5-brane world-volume looks rather similiar to the spectral
curves arising in the exact formulas of the SW anzatz. The way to justify
this proposed in \cite{MMM} was based on parallels with the theory
of integrable systems.

The integrable equations in this context arise as reductions on "unvisible"
dimensions of the equations of motion (better the "square-root" of the: the
BPS-like conditions) of the world-volume theory. In
\cite{Diac} Diaconescu using this idea obtained the Nahm equations. In
\cite{MMM} it was demonstrated that the simplest way for getting
algebraic equation for the topologically nontrivial part of brane
configurations may be searched among the Hitchin systems \cite{Hi}.

The Hitchin system on elliptic curve
\be\label{ell}
y^2 = (x-e_1)(x-e_2)(x-e_3)
\nn \\
x = \wp (z)\ \ \ \ y = 2\wp '(z)\ \ \ \ \  dz \sim 2{dx\over y}
\ee
with $p$ marked points
$z_1,\dots ,z_p$ can be defined by \cite{Ne,LeO} ($i,j=1,\dots,N$)
\be\label{hi}
\bar\partial\Phi _{ij} + (a_i-a_j)\Phi _{ij} =
\sum _{\alpha =1}^p J_{ij}^{(\alpha )}\delta (z - z_{\alpha})
\ee
so that the solution has the form ($a_{ij}\equiv a_i-a_j$)
\be
\Phi _{ij}(z) = \delta _{ij}\left(p_i +
\sum _{\alpha}J_{ii}^{(\alpha )}\partial\log\theta (z-z_{\alpha}|\tau )\right)
+ \left( 1-\delta _{ij}\right) e^{a_{ij}(z-{\bar z})}
\sum _{\alpha}J_{ij}^{(\alpha )}
{\theta (z - z_{\alpha} + {\Im\tau\over\pi}a_{ij})\theta '(0)\over
\theta (z - z_{\alpha})\theta ({\Im\tau\over\pi}a_{ij})}
\ee
The exponential (nonholomorphic) part can be removed by
a gauge transformation
\be\label{laxkri}
\Phi _{ij}(z)\rightarrow (U^{-1}\Phi U)_{ij}(z)
\ee
with $U_{ij} = e^{a_{ij}{\bar z}}$.

The additional conditions to the matrices $J_{ij}^{(\alpha )}$ are
\be
\sum _{\alpha = 1}^p J^{(\alpha )}_{ii}=0
\ee
having the clear meaning that the sum of all residues of a function
$\Phi _{ii}$ is equal to zero, and
\be\label{masses}
\Tr J^{(\alpha )} = m_{\alpha}
\ee
with $m_{\alpha} = {\rm const}$ being some parameters ("masses") of a theory.
The spectral curve equation becomes
\be\label{gencu}
{\cal P}(\lambda ;z) \equiv
\det _{N\times N}\left(\lambda - \Phi (z)\right) = \lambda ^N
+ \sum _{k=1}^N\lambda ^{N-k}f_k(z) = 0
\ee
where $f_k(z)\equiv f_k(x,y)$ are some functions (in general with $k$ poles)
on the elliptic curve (\ref{ell}). If, however, $J^{(\alpha )}$ are restricted
by
\be
\rank J^{(\alpha )} \leq l\ \ \ \ \ \ l<N
\ee
the functions $f_k(z)$ will have poles at $z_1,\dots,z_p$ of the order not
bigger than $l$.
The generating differential, as usual, should be
\be\label{dS1}
dS = \lambda dz
\ee
and its residues in the marked points
$(z_{\alpha}, \lambda ^{(i)}(z_{\alpha}))$ (different $i$ correspond to the
choice of different sheets of the covering surface) are related with the
mass parameters (\ref{masses}) by
\be
m_{\alpha} = \res _{z_{\alpha}}\lambda dz \equiv
\sum _{i=1}^N \res _{\pi _{(i)}^{-1}z_{\alpha}}\lambda ^{(i)}(z)dz
= \res _{z_{\alpha}}\Tr\Phi dz
\ee
It is easy to see that the general form of the curve (\ref{gencu}) coincides
with the general curves proposed in \cite{Witten} at least for
$l=1$, i.e. sourses of rank 1 (for the "rational" case
the torus should be degenerated into a cylinder).

In \cite{MMM} the Toda-chain spectral curve (\ref{suncu}), (\ref{todacur})
has been derived from the $SU(N_c)$ Hitchin system on torus with one marked
point $p=1$ in the double-scaling limit. Of course it is possible to write
down the Hitchin equations directly on the bare cylinder with {\em trivial}
gauge connection
\be
\bar\partial _v \tilde{\cal L}^{TC}(v) = -
\left( e^{-\balpha _0\bphi}E_{\alpha _0} +
\sum _{{\rm simple}\ \alpha}e^{\balpha\bphi}E_{-\alpha}\right)
\delta (P_{\infty}) +
 \left( e^{-\balpha _0\bphi}E_{-\alpha _0} +
\sum _{{\rm simple}\ \alpha}e^{\balpha\bphi}E_{\alpha}\right)
 \delta (P_0)
\ee
where they can be easily solved giving rise to
\be\label{LaxTCHi}
\tilde{\cal L}^{TC}(v) = U^{-1}{\cal L}^{TC}(w)U=
\nn \\
= {\bf pH} + v^{-1}\left( e^{-\balpha _0\bphi}E_{-\alpha _0} +
\sum _{{\rm simple}\ \alpha}e^{\balpha\bphi}E_{\alpha}\right)
+ v\left( e^{-\balpha _0\bphi}E_{\alpha _0} +
\sum _{{\rm simple}\ \alpha}e^{\balpha\bphi}E_{-\alpha}\right)
\ee

\section{Generating Differential}

Let us now turn to more subtle point and discuss how the auxiliary
spectral
Riemann surface is embedded into 11-dimensional target space.
Partially that has been illustrated already above when the explicit formulas
relating 11-dimensional co-ordinates $x^I$ with the internal co-ordinates
on the surface $\lambda $ or $z$ by $\lambda = x^4 + ix^5$ and $z = x^6
+ ix^9$ were presented. In this section I will demonstrate that this
embedding is in fact governed by the generating differential and its
variations.

Already looking at Fig.3 it is clear that the corresponding Riemann
surface is not compact; it means that the metric should have singularities
at "infinities".
Indeed, the metric in the target-space is flat (in case of absense of matter)
$ds^2 \sim \sum _I \left( dx^I\right)^2$ and it means that the area of the
surface is measured by
\be\label{volfor}
\Omega _{\Sigma} \sim \delta z\wedge\delta\bar z + \delta\lambda\wedge
\delta\bar\lambda
\ee
Since the BPS massive spectrum in the theory is determined by
the states corresponding to the 2-branes wrapped over the nontrivial cycles
of "internal" complex manifold -- in our case the Seiberg-Witten
curve -- the BPS masses should be proportional to the area of this surface
\cite{ds}.
This area is measured by another (holomorphic) two form
\be
\Omega \sim \delta\lambda\wedge\delta z
\ee
which is directly related to the variation of generating differential
or to the symplectic form of the corresponding integrable system.

The variation of generating differential over distinguished
subfamily of moduli gives rise to holomorphic differentials
\be\label{var}
\delta _{a_i}dS \sim d\omega_i
\ee
The derivative over moduli (\ref{var}) is taken  after some
connection is chosen -- for example under condition that some function
is covariantly constant. For the differential (\ref{dS}), (\ref{dS1}) the
canonical
procedure implies that the covariantly constant function is $z = \log w$
so that
\be\label{varlambda}
\left.\delta dS\right|_{z = {\rm const}} = \delta\lambda dz =
-\sum _i \delta a_i{{\cal P}_{a_i}\over{\cal P}_{\lambda}}dz
\equiv\sum _i \delta a_i d\omega_i
\ee
From the point of view of $M$-theory one considers an effective
theory on 5-brane world-volume with the co-ordinates $(x^0,\dots,x^3,x^6,x^9)$.
It means that when studying the 4-dimensional effective theory on
"horisontal cylinders" on
Fig.3  one should take the variation of $\lambda $ which has the
sense of vev of some (Higgs) field keeping fixed the world-volume
coordinates $x^6 = \Re z$ and $x^9 = \Im z$. Physically this corresponds
to the fact that we are taking the variation (\ref{var}), (\ref{varlambda})
over the vev's of scalar fields only -- which play the role of physical
moduli in the system.

In principle, this procedure can be correctly defined if one notices that the
differential $dS$ possesses double zeroes
\footnote{More strictly, since only the periods of $dS$ are "observable"
there is always a "canonical" representative in the universality class of the
differentials with fixed periods $dS_1 \cong dS_2$ with double zeroes at
certain points.}; and it is the action of $g$ singular vector fields
$L_{-2}^{(i)}$ at these points that gives rise to the distinguished subfamily
of co-ordinates on the moduli space.
More detailed discussion of the properties of generating differential is
beyond the scope of this note (see the last ref. of \cite{M} for some
details).

%\section{Conclusion}

\section{Acknowledgements}

I am grateful to S.Kharchev, I.Krichever, A.Levin, A.Mironov,
M.Martellini, N.Nekrasov and especially to A.Morozov for valuable discussions
and to T.Sultanov for the help in preparing figures.

The work was partially supported by the
Cariplo Foundation, Institute of Physics of Cambridge University
and RFBR grant 96-02-16117.

\end{document}